\documentclass[showpacs,prl,twocolumn,floatfix,reprint,nobibnotes]{revtex4-2}
\usepackage{amsmath,amssymb,hyperref}
\usepackage{caption}
\usepackage{subcaption}
\usepackage{graphicx} 
\usepackage[font=small,labelfont=bf,
   justification=justified,
   format=plain]{caption} 

\usepackage{blindtext}

\usepackage[
  justification=Justified, 
  calcwidth=.9\linewidth,
]{caption}
\usepackage[ruled,vlined]{algorithm2e} 

\begin{document}

\title{Surfacic Networks}

\author{Marc Barthelemy}
\email{To whom correspondence should be addressed: Email: marc.barthelemy@ipht.fr}
\affiliation{Universit\'e Paris-Saclay, CNRS, CEA, Institut de Physique Th\'{e}orique, 91191, 
	Gif-sur-Yvette, France}
\affiliation{Centre d'Analyse et de Math\'ematique Sociales (CNRS/EHESS) 54 Avenue de Raspail, 75006 Paris, France}

\author{Geoff Boeing}
\affiliation{Department of Urban Planning and Spatial Analysis, Sol Price School of Public Policy, University of Southern California,
  301A Lewis Hall, Los Angeles, CA 90089-0626, USA}

\author{Alain Chiaradia}
\affiliation{Department of Urban Planning and Design, Faculty of Architecture, The University of Hong Kong, Pok Fu Lam, Hong Kong SAR}

\author{Christopher J. Webster}
\affiliation{Faculty of Architecture, and Urban Systems Institute, The University of Hong Kong, Pok Fu Lam, Hong Kong SAR}

\begin{abstract}

Surfacic networks are structures built upon a two-dimensional manifold. Many systems, including transportation networks and various urban networks, fall into this category. The fluctuations of node elevations imply significant deviations from typical plane networks and require specific tools to understand their impact. Here, we present such tools, including lazy paths that minimize elevation differences, graph arduousness which measures the tiring nature of shortest paths, and the excess effort, which characterizes positive elevation variations along shortest paths. We illustrate these measures using toy models of surfacic networks and empirically examine pedestrian networks in selected cities. Specifically, we examine how changes in elevation affect the spatial distribution of betweenness centrality. We also demonstrate that the excess effort follows a non-trivial power law distribution, with an exponent that is not universal, which illustrates that there is a significant probability of encountering steep slopes along shortest paths, regardless of the elevation difference between the starting point and the destination. These findings highlight the significance of elevation fluctuations in shaping network characteristics. Surfacic networks offer a promising framework for comprehensively analyzing and modeling complex systems that are situated on or constrained to a surface environment.

\end{abstract}

\maketitle


\section{Significance statement}
\vspace{-0.3cm}
Networks on non-flat surfaces, such as transportation and urban systems in hilly places, require specialized analyses due to elevation fluctuations. We introduce metrics like lazy paths and graph arduousness to quantify path difficulty and the excess effort which characterizes positive elevation variations along shortest paths. By analyzing simple models and real-life pedestrian networks, we show how elevation variations affect betweenness centrality, path ruggedness, and the overall network efficiency. The additional effort required to travel from one point to another follows a broad distribution, indicating the notable occurrence that many shortest paths entail significant extra exertion to ascend steep slopes. Understanding how elevation fluctuations influence network navigation is essential for analyzing and modeling these systems that are situated on surfaces.

\section{Introduction}


Surfacic networks are defined by a set of nodes and edges that are embedded in a two-dimensional manifold. This manifold could be the plane in the case of usual `plane networks' (i.e. embedded in a plane, but not necessarily planar), or a sphere, or any other rugged surface that correspond to the topography of a place. More generally, this surface could be Earth's surface, biological membranes, or even computational surfaces. As such, they potentially constitute a fundamental concept in various fields, including geography, computer graphics, materials science, and biology. Potential applications could be found in GIS applications where surface networks are used to model transportation networks, utility networks, and other spatially distributed systems on the Earth's surface. It is worth noting here that surfacic networks have to be differentiated from `surface network' in transport planning/geography used for transport networks that are not air or subsurface networks. Note that the elevation of a node could also in principle represent another quantity such as the GDP, average income, etc. and that surfacic networks could be used in some abstract space. Surfacic networks can be considered as a subset of spatial networks \cite{Spatial}. An simple example of such a network is shown in Fig.~\ref{fig:example} (see the part on toy models for details).
\begin{figure}[ht!]
\includegraphics[width=0.45\textwidth]{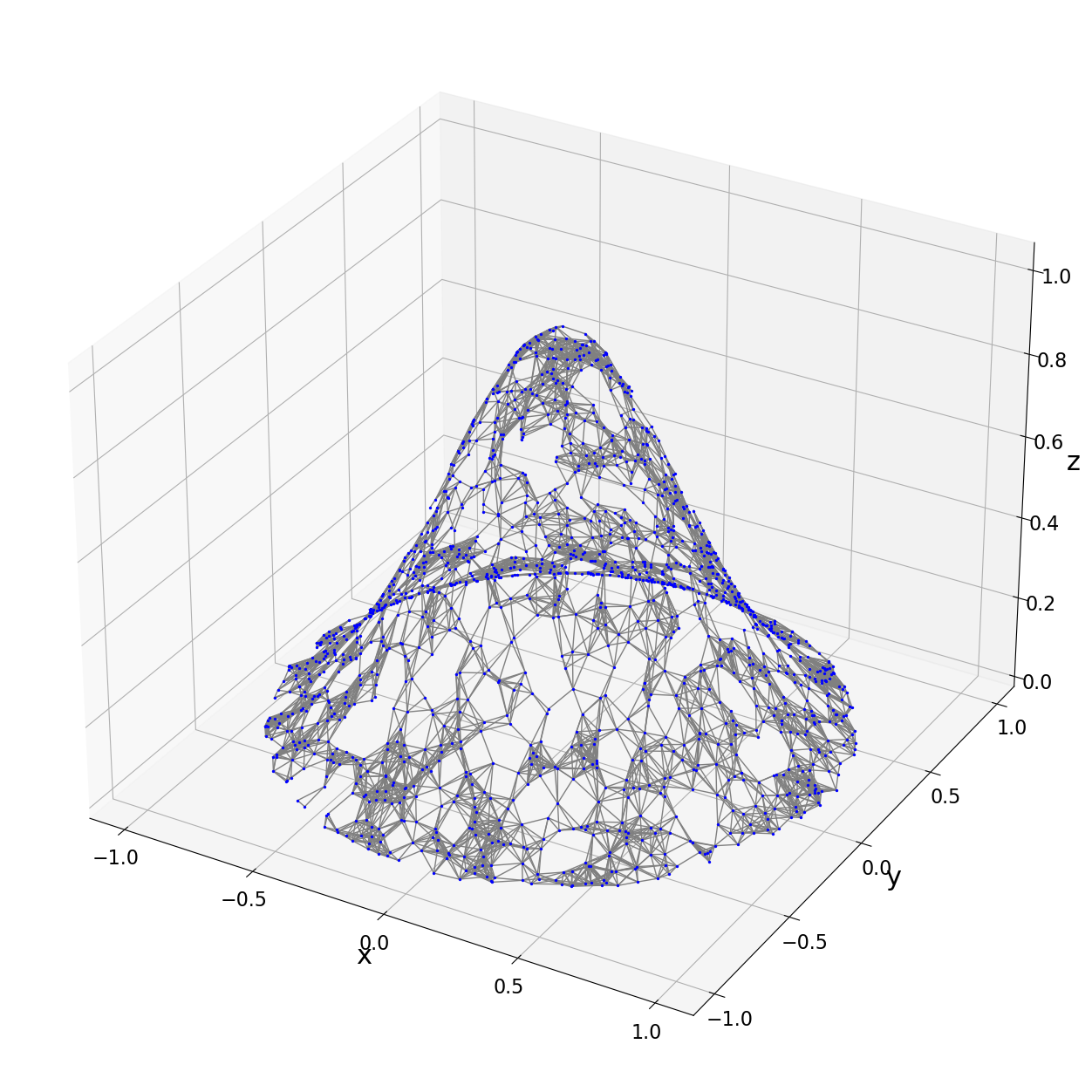}
\captionof{figure}{Example of a surfacic network: random geometric graph on a gaussian surface (the network is constructed here for $N=1500$ points and with threshold $r_0=0.1$, and the gaussian surface is obtained for $z_{\max}=1.0$ and $\sigma=0.4$).}
\label{fig:example}
\end{figure}

Most spatial networks studied so far are embedded in a 2d plane and surfacic networks generalize this by considering non flat two-dimensional manifolds that can have curvature fluctuations. Other spatial networks could be embedded in 3d networks (such as the important case of the neuronal network \cite{Bullmore,Sporns}) and could be defined as `volumetric' networks or `physical' networks \cite{Dehmamy}. We therefore have the following nested inclusion between these sets of networks:
\begin{align}
  \nonumber
  \mathrm{Surfacic} \subset\mathrm{Volumetric}\subset\mathrm{Spatial}\subset\mathrm{Networks}
\end{align}

We mention here the existence of the term `surface networks' that was quoted in \cite{surf} where the authors study data-driven representations for three-dimensional triangle meshes, which are one of the prevalent objects used to represent 3D geometry.

The geometry of the surface will influence the structure and the behavior of a surfacic network. Geometric properties such as curvature, topology, and spatial constraints certainly play a crucial role in shaping the connectivity and dynamics of these networks. This is essentially due to spatial constraints imposed by the surface on which the networks are embedded. These constraints affect the arrangement of nodes and edges, as well as the navigation and flow of information or resources within the network. In contrast to spatial networks that have been thoroughly studied, the impact of the embedding topography on the network structure has not been systematically considered in the literature. There are a few exceptions such as \cite{Sun,Zhang, Cooper} that considered the impact of elevation fluctuations on pedestrian paths. Other urban networks are naturally impacted by elevation fluctuations, such as the water distribution network where node elevation is an important information that has to be taken into account \cite{Wannapop}. At a more theoretical level, network geometry studies (see \cite{Boguna} and references therein) are probably connected to surfacic networks. However, further studies are needed in order to exploit this possible correspondence.

A practical example is the road network that follows the topography of a city. This type of networks was extensively studied (see for example \cite{Xie,Strano,Marshall,Boeing2021,Spatial} and references therein), and in this case, maps represent in general a vertical projection (`from above') of the surfacic network, which can be very misleading (in particular in terms of the physical effort for a pedestrian for example). This type of spatial network is different from 3d spatial networks (such as the brain for example) and constitute a prime example of a surfacic network. Some specific measures will then be needed to characterize the importance of the `third' dimension. The example of road and street networks is particularly relevant for pedestrians and perhaps even more so for cyclists, where the altitude variation represents an effort for individuals. These networks are fundamental for analyzing spatial relationships, optimizing routes, and supporting decision-making in urban planning, logistics, and environmental management. 

In order to quantity the impact of the surface shape on the network structure, we will introduce a set of new tools that take into account the third dimension described by the elevation of nodes. In addition to an adjacency matrix, each node is described by its coordinates $(x,y,z)$ where the elevation (or height) $z$ can display large fluctuations. In the general case of volumetric networks the elevation $z$ has no constraint, while for surfacic networks, the elevation $z$ is a function of the coordinates: $z=F(x,y)$. The function $F$ defines the two-dimensional manifold embedded in 3d space. In the constant case $z=\mathrm{const.}$, we recover usual plane networks. 

There are many important theoretical questions about surfacic networks. For example, we need to understand what is the impact of elevation fluctuations on the usual properties of graphs. This concerns the spatial distribution of the betweenness centrality, the shape of shortest paths, etc. For this we can monitor on toy models the evolution from a graph on a flat plane to a graph on a deformed surface. Also, it is important to understand the impact of elevation fluctuations on standard graphs such as the minimum spanning tree and other benchmarks. 

We introduce a toy model where the network is constructed over a paraboloid which will allow us to discuss various properties related to shortest paths, or to the betweenness centrality. We then consider empirical examples of road and pedestrian networks in real cities, for which elevation differences are particularly relevant.

\section{Tools and measures}

The coordinates of a node $i$ is denoted by $x_i.y_i,z_i$. The euclidean distance between two adjacent nodes can be generalized under the form
\begin{align}
\ell_{\gamma}(i,j)=\sqrt{(x_i-x_j)^2+(y_i-y_j)^2+\gamma (z_i-z_j)^2}
\label{eq:dist}
\end{align}
With this definition, we can monitor the influence of elevation: for $\gamma=0$ the network is `flat' and for $\gamma=1$ we consider the full impact of elevation fluctuations. 

We note that surfacic networks can be seen as weighted 2d networks where nodes are weighted by their elevation, but encoding the elevation in
weights could be very complex. In this respect the network could be seen as a fitness or hidden variable model \cite{Calda:2002, Boguna:2003}. An important fact here is that the weights (elevation) are non-trivially correlated (depending on the surface) and depend in a complicated way on the shape of the surface. In the case where the surface is defined by a function of the form $z=F(x,y)$ (and the network connects the nodes of coordinates $(x_i,y_i,  z_i=F(x_i,y_i)$), we can construct the weights that encodes the elevation. For example, if an edge $e$ connects two nodes
  $M(x,y,z)$ and $M'(x',y',z')$, and if we assume that these nodes are
  close to each other, the weight that represents the total distance
  is given (if one assumes that the surface is differentiable):
  \begin{align}
    w(e)=\sqrt{
    \Delta x^2+\Delta   y^2+ (\overrightarrow{\nabla}F\cdot\overrightarrow{u_e})^2
    }
  \end{align}
  
  where $\Delta x = x'-x, \;\Delta y=y'-y$, and $\overrightarrow{u_e}=(\Delta x,\Delta y)$. We see here that, although the representations in terms of a weighted 2d graph and a surfacic network are equivalent, the formulation with weights could rapidly become very complex.

Also, in this  representation, we see that if the network is constructed on a  differentiable
surface $z=F(x,y)$, if $xi \approx xj$ and $yi \approx yj$ we have: $zi \approx zj$.
We expect such a continuity property to be valid in general for real-world networks (such as road networks for example), but there could be exceptions such as in the case of pedestrian networks, where elevators for example induce discontinuities.

\subsection{Excess effort}

A shortest path $SP(i,j)$ from $i$ to $j$ incurs a total elevation difference given by
\begin{align}
	\sum_{e\in SP(i,j)}\Delta z(e) = z(j)-z(i)\equiv \Delta z(i,j)
\end{align}
where $\Delta z(e)$ is the elevation difference of edge $e$ (in the direction of the shortest path $\Delta z(e)=z(target)-z(source)$).
Any continuous path must satisfy this equation, but if there is a discontinuity (such as
an elevator for example) we would have to include vertical edges in order to still satisfy it.

In a human navigation domain, where positive slope requires more effort, we define
the elevation effort between $i$ and $j$ as
\begin{align}
	\Delta z^+(i,j)=\sum_{e\in SP(i,j)}\Delta z(e)\theta(\Delta z)
\end{align}
where $\theta(x)$ is the Heaviside function (we can similarly define $\Delta z ^-$ for negative slope segments to represent down-hill energy, time, etc. advantage). 

If the shortest path from $i$ to $j$ is monotonically increasing, we have $\Delta z^+(i,j)=\Delta z(i,j)$. However in real cases, we have to go up and down so that we have to climb up a distance $\Delta z^+$ larger than the elevation difference $\Delta z$. There is therefore an excess effort $E(i,j)$ that we can measure with 
\begin{align}
E(i,j)=\frac{\Delta z^+(i,j)}{\Delta z(i,j)}-1  
\end{align}
With these measures, we can then study various network statistics such as its average, distribution, etc.

\subsection{Lazy paths and graph arduousness}

In general the shortest path on a spatial network is computed using as the weight the euclidean length of a link: $w(e)=\ell_{1}(e)=\sqrt{\Delta x^2+\Delta y^2+\Delta z^2}$. The (weighted) shortest path is then the one that minimizes the sum of $w(e)$ along it. 

In surfacic networks, however the elevation of a node is important. We can therefore define shortest paths so that they weight elevation difference, rather than minimizing total distance only. In particular for pedestrians it makes sense to avoid paths with a large (positive) elevation difference. We therefore assign the following weight to an edge
\begin{align}
w(e)=\begin{cases}\ell_1(e)+\mu\Delta z(e)\;\;&\mathrm{if}\;\;\Delta z(e)>0\\
\ell_1(e)\;\;&\mathrm{if}\;\;\Delta z(e)<0
\end{cases}
\label{eq:w}
\end{align}
where $\Delta z(e)=z(\mathrm{end\; node})-z(\mathrm{starting\; node})$ and where $\mu>0$ is a parameter. We then look for the optimal path that
connects two nodes such that the total weight $W=\sum_{e\in path}w(e)$ is minimum. We note that with this choice of weight, smoother slopes are always favored. However, in reality, a pedestrian might prefer a path with a steeper slope followed by a longer flat section (see SI Fig.~S1 in the Supplementary Information). It would be interesting to explore more complex weight functions that account for this behavior.

The parameter $\mu$ governs the relative weight of the elevation effort and the
length of the trip. For $\mu=0$, this optimal path is the usual shortest path that minimizes the total distance.
When $\mu\gg 1$, the optimal path essentially minimizes the excess effort.   The
  choice $\mu=1$ corresponds then to the case where we choose a longer
  path if the detour is of the order (or smaller) than the excess
  effort difference. More precisely, assume there is a path 1 characterized by a
  total   weight $W_1=L_1+\mu\Delta z_1$ and another path 2 by
  $W_2=L_2+\mu\Delta z_2$. Assume that $L_1>L_2$. If $\Delta
  z_1>\Delta z_2$, there is no ambiguity the shortest path has also
  the smallest excess effort and the optimal path is $L_2$. In
  contrast, if $\Delta z_1<\Delta z_2$, we are in an ambiguous case:
  the path $L_1$ is longer but it has a smaller excess effort. The
  resulting optimal path will then depend on the value of $\mu$, and
  the path $1$ will be chosen if $W_1<W_2$ which implies that
  $L_1-L_2<\mu (\Delta z_2-\Delta z_1)$ which indeed corresponds to
  the fact that what we loose in detour, we gain in excess
  effort. Larger values of $\mu$ would give more weight to the
  elevation effort. We show in the SI (see Fig. S2) the evolution of the arduousness
  for different values of $\mu$. Here and in the following, we will use $\mu=1$.
 
  We thus look for the path that minimizes the total weight with weights given by Eq.~\ref{eq:w}. The optimal path is then the shortest one that minimizes the total elevation effort, which we coin `lazy path', and we denote its length by $L_{lazy}(i,j)$  which is given by
\begin{align}
  L_{lazy}(i,j)=\sum_{e\in LP(i,j)}\ell_1(e)
  \label{eq:lazy}
\end{align}
where $LP(i,j)$ is the set of links belonging to the lazy path from $i$ to $j$. We leave $\Delta z <0$ unweighted in this experiment, but note that we could assign a negative weight to represent any advantage conferred by a `down-slope' link. In pedestrian networks a negative $\Delta z$ confers energy, time and psychological advantage in navigating a surfacic network. This may not be the same in other domains. Note that this lazy distance naturally induces directionality in the network: $L_{lazy}(i,j)\neq L_{lazy}(j,i)$. 

We denote by $L_{tot}(i,j)$ the total length of the shortest path between $i$ and $j$, which minimizes the following expression
\begin{align}
  L_{tot}(i,j)=\sum_{e\in SP(i,j)}\ell_1(e)
   \label{eq:sp}
\end{align}
where $SP(i,j)$ is the set of edges of the shortest path between $i$ and $j$. Note that the expressions in Eqs.~\ref{eq:lazy} and \ref{eq:sp} both represent a total length.
However, for Eq.~\ref{eq:lazy}, it corresponds to the total length of the path that minimizes $\sum_e w(e)$, while for Eq.~\ref{eq:sp}, the path minimizes $\sum_e \ell_1(e)$.

We then construct the ratio of these two lengths
\begin{align}
	A(i,j)=\frac{L_{lazy}(i,j)}{L_{tot}(i,j)}-1
\end{align}
For a flat network, we have $A(i,j)=0$. If for a pair $(i,j)$, we have a large value of $A$, it means that the lazy path is much longer than the shortest path and that elevation is critical. We can average this quantity $A(i,j)$ over all pairs of nodes and obtain the average `arduousness' of the graph $G$
\begin{align}
	A(G)=\frac{1}{N(N-1)}\sum_{i\neq j}A(i,j)
\label{eq:ardu}	
\end{align}
The arduousness of a flat network is defined as $A(\text{flat}) = 0$. Higher values of $A$ indicate greater significance of elevation,
implying that, without a detour, one must contend with substantial elevation differences. We note that in other fields such as
optimization on non-convex problems, we could imagine that arduousness might provide a measure of search difficulty in data with many local optima.

\section{Toy models}

We introduce the concept of surfacic networks using pedestrian paths through a topographical city, but there are certainly many models of surfacic networks that can be imagined. Potentially, any network model can be embedded on any surface, leading to an infinite number of possible models. Here, we explore three simple models that allows to investigate different aspects of surfacic networks. First, we consider a network constructed over a paraboloid, mimicking cases where the topography has a single peak. The main parameter is then the height of the peak (rescaled by the typical size of the area) and we can discuss various properties when this maximum height is varied. In the second `gaussian' model, the width of the peak can also be monitored. Finally, we consider another random null model where the elevation of a node is a random variable. This allows us to investigate the impact of elevation fluctuations on various aspects of surfacic networks.

For practical purposes and expositional simplicity, we construct the toy networks as a function of the underlying topography. We could have merely layered a network topology on a topographical surface.

\subsection{The parabolic model}

The idea of this model is to mimic a network that is defined on a surface which has one main `peak'. We thus assume that the $x_i,y_i$ coordinates of the nodes are 
random distributed in the plane (typically in the disk of radius $R=1$), and that for each node $i$, the elevation $z_i$ is given as a function of their (random) planar coordinates $x_i,y_i$ by
\begin{align}
z_i=z_{max}(1-x_i^2-y_i^2)
\end{align}
More generally, for a surface defined by $z=f(x,y)$, we would choose $z_i=f(x_i,y_i)$ with random $x_i$ and $y_i$. We can then choose any rule to construct the network on top of this surface. Here, we choose to construct a random geometric graph (RGG) \cite{Gilbert:1961} where two nodes $i$ and $j$ are connected only if their distance is less than a threshold $\ell_1(i,j)\leq r_0$. We choose $r_0$ such that the average degree is $6$ ensuring the existence of a large giant component (see \cite{Spatial} and references therein). We note here that Krioukov et al. \cite{Krioukov} considered a RGG built on a hyperbolic geometry, a first example of a surfacic network (see \cite{Duchemin} for a recent review on RGG). 

In this simple case considered here, we thus have a paraboloid embedding the network which allows us to investigate the impact of elevation by varying $z_{max}$ from $0$ at the minimum (which corresponds to a flat network) to $z_{max}$ at the maximum, which is much larger than the typical size $R$. For this model, we will discuss its arduousness, excess effort, and the spatial distribution of betweenness centrality. 
\begin{figure*}[ht!]
  \includegraphics[angle=0, width=\linewidth]{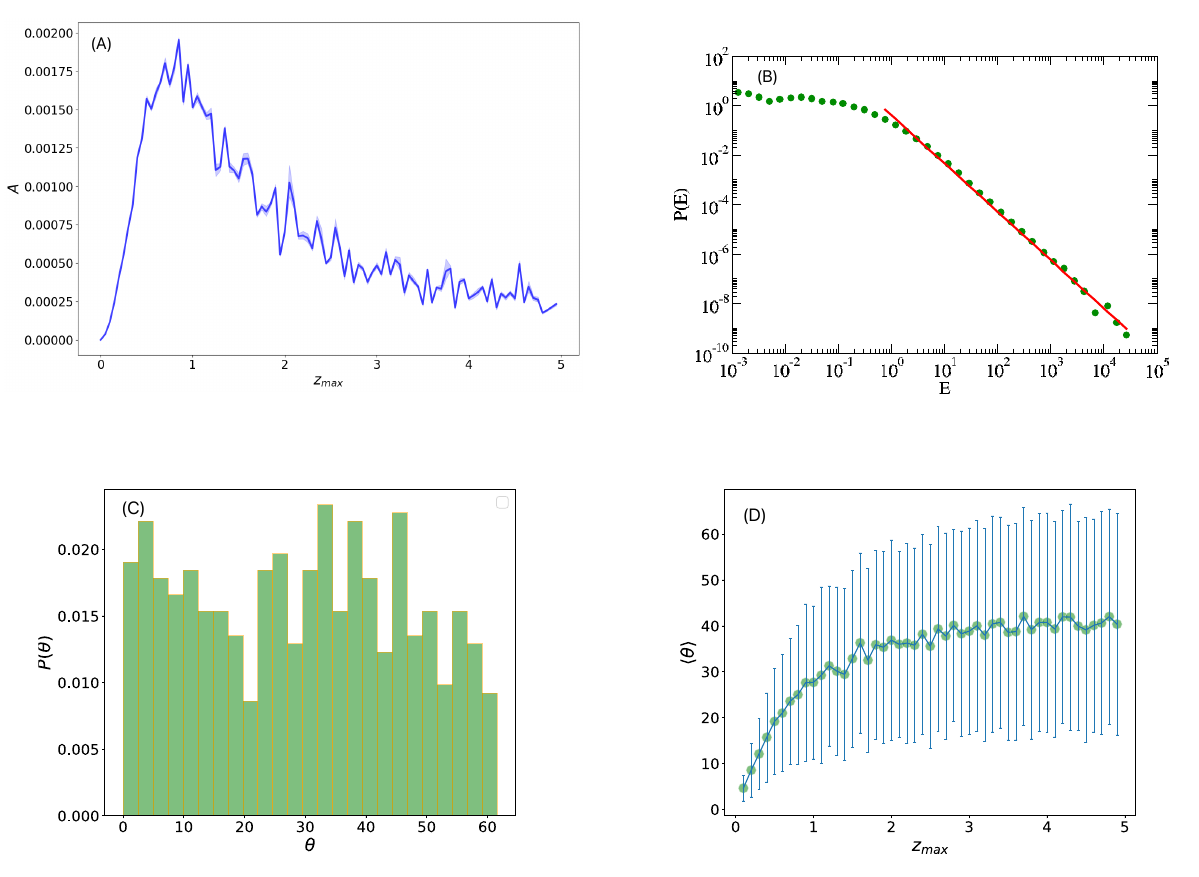}
\caption{The parabolic model. (A) Arduousness of the graph computed for the parabolic model when $z_{max}$ is varied. (B) Distribution $P(E)$ of the excess effort (for $z_{\max}=1$, $N=200$ and $1000$ configurations). The straight line is a power law on the tail with exponent $2.0$ ($r^2=0.99$).
  (C) Distribution $P(\theta)$ of the angle of edges ($N=200$, $1000$ configurations, $z_{\max}=1$). (D) Average slope versus $z_{\max}$ ($N=200$, $1000$ configurations).}
\label{fig:arduous}
\end{figure*}

\subsubsection{Arduousness}

We compute the arduousness (Eq.~\ref{eq:ardu}) for the parabolic graph when $z_{max}$ varies from $0$ (flat network) to $z_{max}$ large (in our experiment, we choose $z_{max}=5$). We obtain the result shown in Fig.~\ref{fig:arduous}(A). We observe on this figure the presence of a maximum (and we expect this feature to be quite general): when $z_{max}$ is small, the arduousness is small as they are many shortest paths that don't necessitate to climb large elevation differences. For very large $z_{max}$ most shortest paths avoid the peak of the mountain and the lazy and shortest paths are very similar. There is therefore a maximum value $z_{max}=z^*$ for which many shortest paths actually go through the peak and entail climbing many edges with a relatively large $\Delta z$. Around this point $z^*$ there is a large difference between shortest and lazy paths.

\subsubsection{Excess effort}

We compute the average excess effort $\langle E\rangle$, and obtain the distribution $P(E)$ shown in Fig.~\ref{fig:arduous}(B). We observe on this figure
that this distribution has a power law tail of the form $P(E)\sim E^{-\gamma}$ with $\gamma\approx 2$. This behavior indicates that there is a small but non negligible probability of finding a pair of nodes such that $E$ is very large (up to $10^4$). Large values of $E$ typically occur when the distance $\ell_1(i,j)$ is large (i.e., nodes $i$ and $j$ are far apart) while their elevation difference is small. In such cases, the shortest path traverses nodes with significant elevation changes, leading to a large $\Delta z^+$, and consequently, a very high value of $E$.

We also measure the distribution of the angle $\theta$ of edges (see Fig.~\ref{fig:arduous}(C)). The angle for an edge $e$ connecting nodes $M=(x,y,z)$ and $M'=(x',y',z')$ is defined as $\theta=\mathrm{atan} (z'-z)/[(x'-x)^2+(y'-y)^2]^{1/2}$. It is here an indication of the slopes experienced by shortest paths. In the case
of the case of the parabolic model find a roughly uniform distribution. The average angle $\langle\theta\rangle$ of edges depends however on $z_{\max}$ and we show the result on Fig.~\ref{fig:arduous}(D). We observe a rapid increase when $z_{\max}$ goes to 1, followed by a plateau for a value $\langle\theta\rangle\approx 40$ degrees. It is interesting that $\langle\theta\rangle$ never crosses this value: when $z_{\max}$ becomes too large, shortest paths naturally avoid the steep edges on paths that are also too long. 


\subsubsection{The spatial distribution of the BC}

Betweenness centrality (BC) is an important measure in networks and represents a reasonable proxy for the traffic on a link (see for example \cite{Spatial}). It points to structurally important nodes that can be considered as bottlenecks in a network. For node $i$, it is defined as 
\begin{align}
g(i)=\frac{1}{N(N-1)}\sum_{s\neq t}\frac{\sigma_{st}(i)}{\sigma_{st}}
\end{align}
where $\sigma_{st}$ is the number of (weighted) shortest paths between $s$ and $t$ and 
$\sigma_{st}(i)$ is the number of (weighted) shortest paths between $s$ and $t$ that go through the node $i$ ($N$ is the number of nodes in the graph). In this context, shortest paths are determined based on a specific weight (referred to as weighted BC by some authors, but here simply as BC), where we consider the length of an edge as its weight.

For most spatial networks, there is some correlation (on average) between betweenness centrality and distance to the center \cite{Spatial}. For example, on a regular 1d lattice of length $L$, betweenness centrality is $g(x)=x(L-x)$ for $x\in [0,L]$, and has a maximum at $L/2$ (note that for a disordered planar network, the BC can be more complicated, see for example \cite{Lammer,Spatial}). We expect that elevation will be an important factor governing the spatial distribution of the BC and it is interesting to study the correlation between the BC and the elevation. We note that some analysis of flows on a pedestrian surfacic network, can be found in \cite{Zhang} where the authors test the association between various BC metrics and pedestrian flow counts. 

\begin{figure}[h!]
        \includegraphics[width=0.45\textwidth]{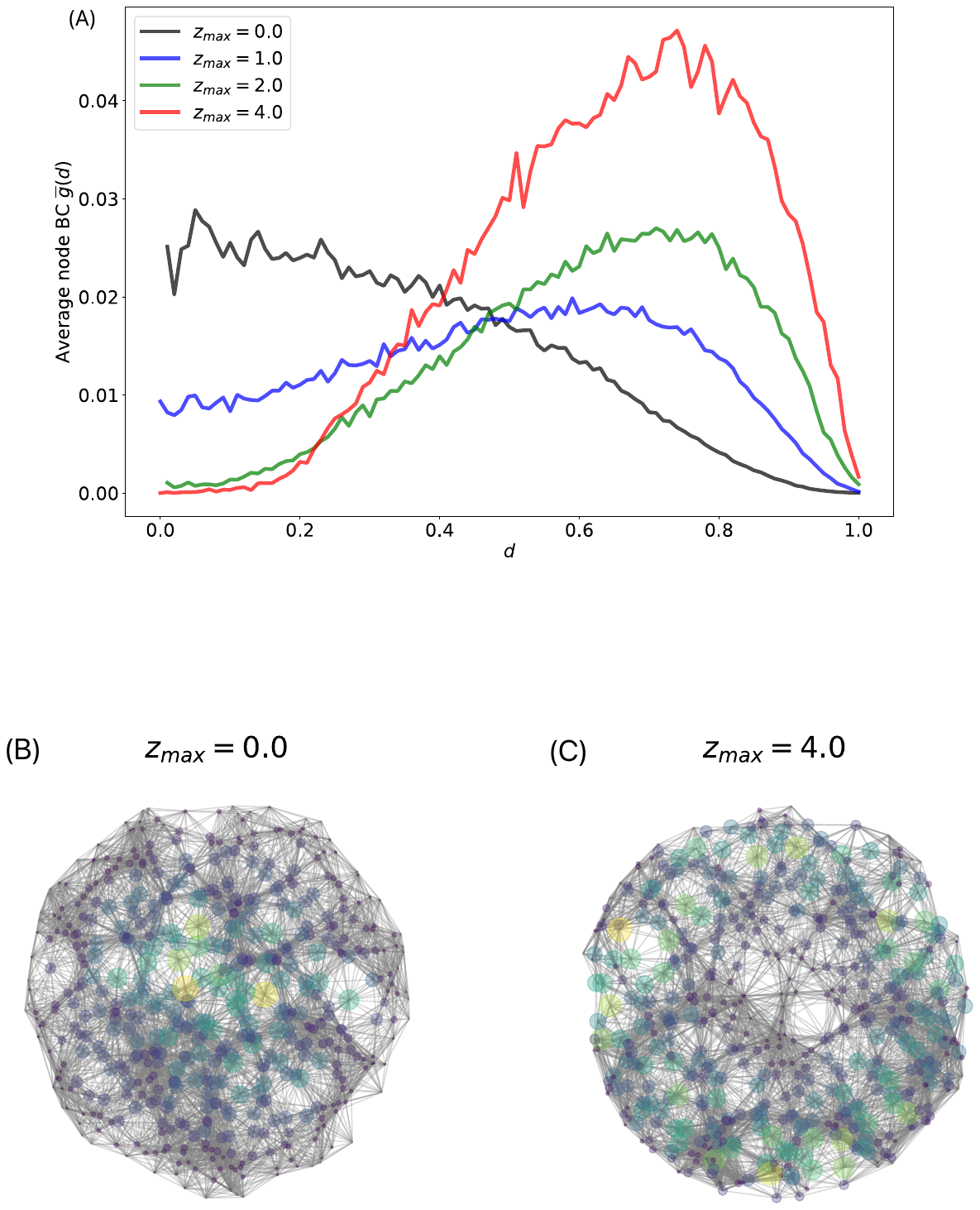}
	\caption{(A) Spatial distribution of the BC for parabolic networks: we plot here the average BC versus the distance to the center for different values of $z_{max}$ (computed for $100$ nodes and averaged over $100$ configurations). (B,C) Spatial distribution for BC for a flat network (left, $z_{max}=0.0$) and a non-flat network with a maximum at the center (right, $z_{max}=4.0$). The color code and size of nodes is proportional to the BC (Calculation are done on random geometric graphs of size $N=100$, $200$ configurations, and average degree $\langle k\rangle=12$).}
	\label{fig:BC}
\end{figure}

For flat (isotropic) networks, BC decreases with the distance $d$ from the center. We can then plot the average BC versus the distance $d$ for different values of $z_{max}$ (the peak of the parabolic surface). The result is shown in Fig.~\ref{fig:BC}(top). 
For small $z_{max}$ we recover the usual flat network behavior: high BC nodes are close to the center of the graph (which corresponds to the center of the disk here). When $z_{max}$ increases, we observe that there is a crossover to another regime where the large BC nodes are actually close to the boundary where the elevation is small. This can be confirmed by visual inspection on the two cases presented in Fig.~\ref{fig:BC}(bottom) (this figure is shown for a random geometric graph with a large average degree; for smaller value of $\langle k\rangle$ the phenomenon exists but with a smaller amplitude). 

We provide a simple hand-waving argument to estimate when this crossover happens (for a disk of radius $R=1$). We consider two diametrically opposed nodes $A$ and $B$ with both having a small elevation and located at a radial distance of order $\sim R=1$ from the center. The length $\ell_S(A,B)$ of the path going through the center (ie. the `peak' of the mountain) is given by the arc length of the parabola from $-R$ to $+R$ which is given by
\begin{align}
\nonumber
  \ell_S(A,B)&= 2\int_0^R\sqrt{1+(2z_{max}\tau)^2}\mathrm{d}\tau\\
  \nonumber
  &= \sqrt{1+4z_{max}^2}+\\
&\frac{1}{2z_{max}}\log\left[2z_{max}+\sqrt{1+4z_{max}^2}\right]
\end{align}
The equatorial path (i.e. that avoids the mountain peak and goes around) is of length $\ell_E(A,B)=\pi R=\pi$. For $z_{max}$ small, it is shorter to go straight from $A$ to $B$: $\ell_S<\ell_E$. When $z_{max}$ increases $\ell_S$ increases and at a certain point $\ell_S$ and $\ell_E$ become equal. This happens for $z_{max}=z^*\approx 1$. Above this value $z^*$, it becomes more optimal to avoid the peak of the mountain and in this case the small elevation nodes become central.

\subsection{The Gaussian model}

\begin{figure}[ht!]
\includegraphics[width=0.4\textwidth]{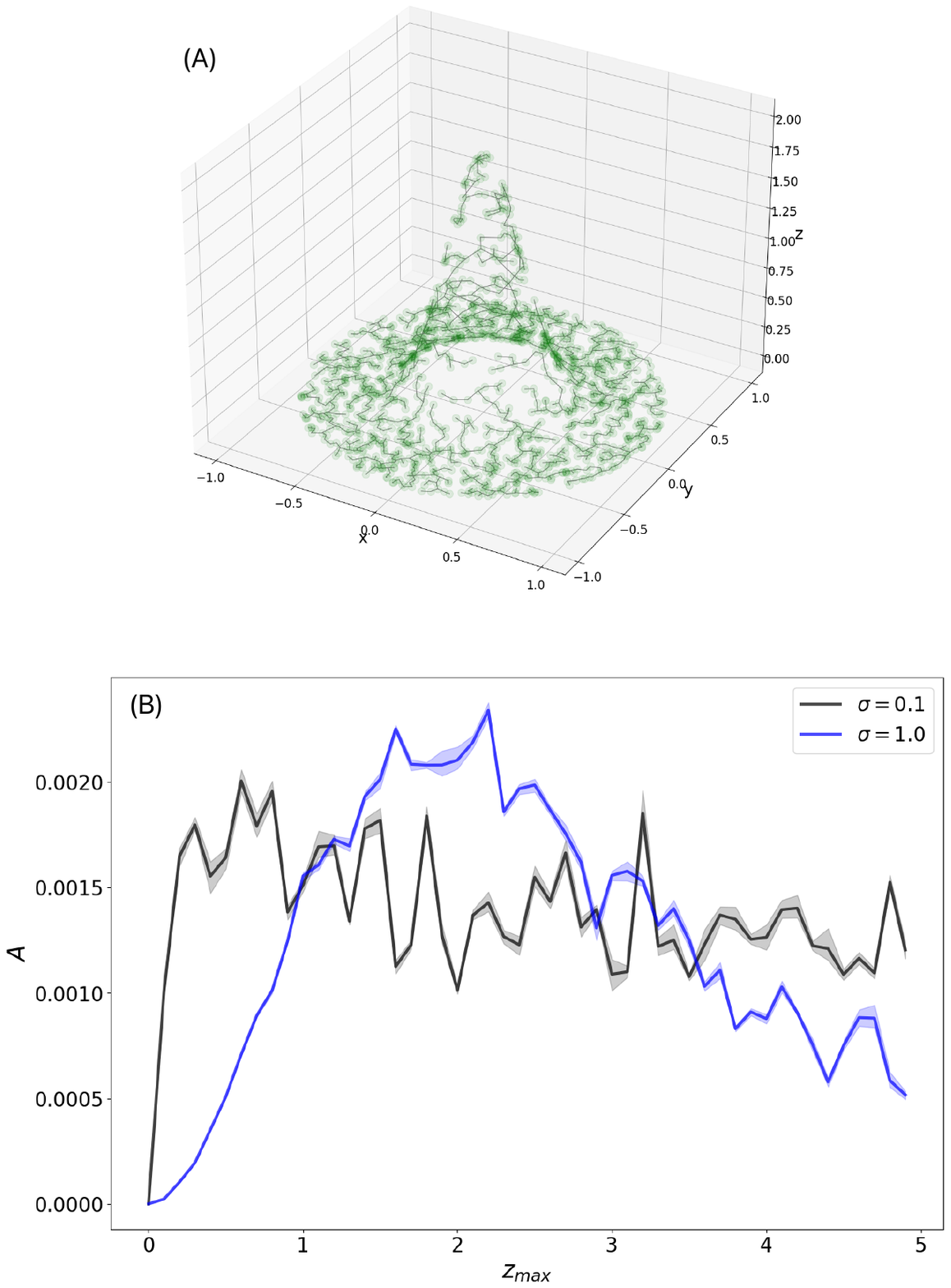}
		\caption{(A) Example of a gaussian shaped surfacic network (here an  euclidean minimum spanning tree for $N=1500$ nodes, and constructed over a gaussian surface with  $z_{\max}=2/0$, and $\sigma=0.2$). (B) Arduousness computed for the gaussian model when $z_{\max}$ is varied and for two different values of $\sigma=0.1$ and $1$ (results are computed for $N=200$ nodes, $100$ configurations, and for $100\times 100$ pairs of nodes).}
	\label{fig:arduousness_gaussian}
\end{figure}

Next we experiment with a gaussian shaped surface (see an example in Fig.~\ref{fig:arduousness_gaussian}. As in the parabolic case, we choose random coordinates $(x_i,y_i)$ and compute the surfacic network as a function of an underlying topography
\begin{align}
z_i=z_{max}\mathrm{e}^{-(x_i^2+y_i^2)/2\sigma^2}
\end{align}
As in the parabolic case, we can monitor the effect of the peak height $z_{\max}$ but also the width of this peak with $\sigma$. For a small $\sigma$ the peak is very steep, while for a large $\sigma$ (compared to the system size $R$), the slope of paths towards the peak of the mountain are shallow. 

We compute the arduousness in this gaussian case for two different values of $\sigma=0.1$ and $\sigma=1.0$. The results are shown in Fig.~\ref{fig:arduousness_gaussian}.
We observe that for $\sigma=1$ we recover a behavior similar to the one obtained in the parabolic case. However, when $\sigma$ is too small (here $\sigma=0.1$), the peak is very narrow and the influence of $z_{\max}$ is limited as few shortest paths actually go over it (and go around it), leading to an almost constant arduousness.

\subsection{The random model}

A simple model for studying the impact of fluctuations on various properties of surfacic networks can be defined as follows. We start from a flat graph $G$ where nodes have coordinates $(x_i,y_i)$ and we assign to each node $i$ a random elevation $z_i$ defined as
\begin{align}
	z_i=\overline{z}+\sigma\xi_i
\end{align}
where $\xi_i$ is a random number of order $1$, and $\sigma$ determines the scale of the fluctuations. We simulated this model for a uniformly distributed random set of points. Here we focus on just two different aspects: the structure of shortest paths and the minimum spanning tree.

\begin{figure}[ht!]
        \includegraphics[width=0.45\textwidth]{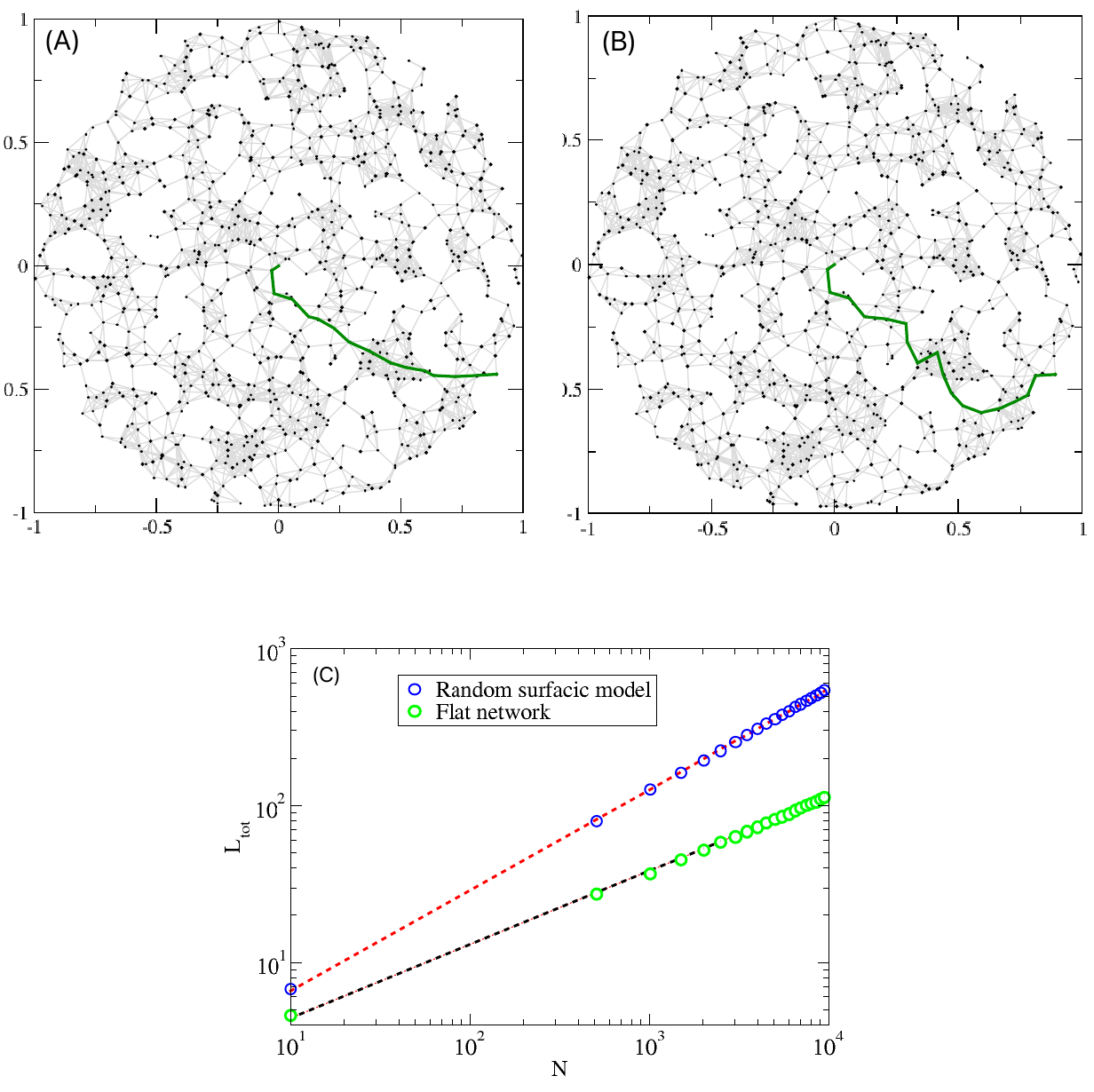}
		\caption{(A,B) Example of the deformation of a shortest path. On the left (A), we show the shortest path computed for the flat graph, and on the right (B) we show the shortest (between the same pair of nodes) computed for the surfacic network. (C) Scaling of the total length with the number of nodes. Dotted lines are power law fits of the form $L_{tot}\sim N^\tau$. For the flat network, we obtain $\tau\approx 0.46$ (in agreement with the standard result $\sqrt{N}$ and for the random model, we obtain $\tau\approx 0.64$ a three-dimensional behavior (with $r^2=0.99$ in both cases). These results are obtained for an average over $100$ configurations and the random surfacic network is obtained for $\sigma=1$.}
	\label{fig:compareSP}
\end{figure}

With this model, we can observe how elevation fluctuations alter shortest paths. This is important from a practical point of view for city maps: if the shortest paths are very different, it means that we should be careful when using the view from above (or the projection of the map) for navigating in the city. We tested this on a random geometric graph (see \cite{Dettmann,Spatial} and references therein) constructed over a set of points in the two dimensional plane and where we assign to each point $(x_i,y_i)$, a random elevation $z_i$. Results are shown in Fig.~\ref{fig:compareSP}(top). We have highlighted in this figure a pair of nodes and the corresponding shortest path. On the figure left, we consider the flat graph, and in the figure on the right, we take into account the (random) elevation. We observe that in the presence of elevation, the shortest path is very different and typically avoids large elevation nodes, as expected from the BC analysis.
%

There are many other aspects that we could study. In particular, we could study the euclidean minimum spanning tree (EMST) for this model and characterize the importance of elevation fluctuations in the structure of the EMST (see SI Fig.~S3 that highlights the impact of elevation on the EMST structure). It is well known \cite{Beard} that for flat networks, the total length $L_{tot}$ of an EMST scales as $L_{tot}\sim N^\tau$ with $\tau=1/2$. Results shown in Fig.~\ref{fig:compareSP} show that for the surfacic model with $z_i=\sigma u_i$ (where $u_i$ is a random number in $[-1,1]$), the exponent is different $\tau\approx 0.64$ (for $\sigma=1$). For a dimensional network, we expect that the typical distance between nodes is of order $1/N^{1/d}$ which leads to a total length scaling as $N^{1-1/d}$. The result obtained here thus corresponds to $d=3$, showing that the fluctuations in this model are enough to destroy the surfacic feature of the network which now ressembles more a three dimensional network. Real-world surfaces doesn't display this sort of large fluctuations, but we believe that the scaling of the MST on surfaces deserves probably further study, with possible crossovers for some models from a 2d to a 3d behavior.

\section{Empirical analysis}

Pedestrian (and road) networks are typical examples of surfacic networks where elevation plays a critical role. Elevation directly influences the accessibility of various urban points, shaping numerous aspects of a city's spatial economy, including land values, the viability of commercial centers, and the balance between jobs and housing. Despite its importance, research on pedestrian networks remains less developed than that on road networks \cite{Rhoads}, even though it is a vital component of urban infrastructure. Notably, elevation affects pedestrian speeds, influencing the geometry of minimum-time paths. Early discussions on footpaths in hilly terrains were initiated in \cite{Tobler:1993} and later revisited in \cite{Goodchild:2020}. For flat cities (i.e. with small fluctuations of the elevation - for example, the maximum elevation for Paris is $130$ meters), elevation difference is irrelevant. This is in contrast with other cities constructed over hilly surfaces. This is for example the case of Hong Kong island, a densely inhabitated urban space rising from sea-level to over 500 meters and having very many steeply sloped roads and paths. We will also consider the case of San Francisco, which is interesting in the sense that most hilly cities have curving streets to accommodate the topography, but San Francisco is an outlier case where the road grid was laid over the top of the hills, so some parts of the city have particularly steep streets. These two examples will help us to illustrate our measures and results. These networks are extracted from the extensive dataset provided in \cite{Boeing2021}. More specifically, geopackages are provided for each city, and the graph was topologically simplified such that nodes represent dead-ends and junctions, but full correct edge geometry is maintained. This graph's edges contain attributes for `length' (representing Haversine distances between original unsimplified nodes, then summed when graph was simplified), and also a 3d length that represents 3d euclidean distances between projected original unsimplified nodes, then summed when graph was simplified (see the SI for more details about the network construction and the corresponding pseudocode).

\subsection{Elevation and slope distributions, excess effort}

Simple statistics for pedestrian networks can be measured using
the extensive dataset provided in \cite{Boeing2021,Boeing2017} such as the elevation distribution, fluctuations across cities in the world, etc. (see SI Fig.~S4 for the elevation distribution for cities in different countries). Beside standard statistical indicators, we compute for a given city the average elevation of nodes and their Gini coefficient: if all the elevations are (almost) equal the Gini coefficient is (close to) zero and if a few nodes have a very different elevation than the rest, the Gini coefficient will be close to one. The result is shown for a few countries in the SI Figure S5. We observe that the average elevation $\overline{z}$ decreases with the Gini coefficient $G$: cities with large fluctuations are found at a lower altitude (a rough fit gives $\overline{z}\sim 1/G$). This might seem counter-intuitive but is because, as with the case of Hong Kong island, coastal and deltaic cities often spread into surrounding hilly terrain. On the other hand, high altitude cities are often built on mountain plateaux (such as Mexico City).
 
\begin{figure*}
	\centering
        \includegraphics[width=\textwidth]{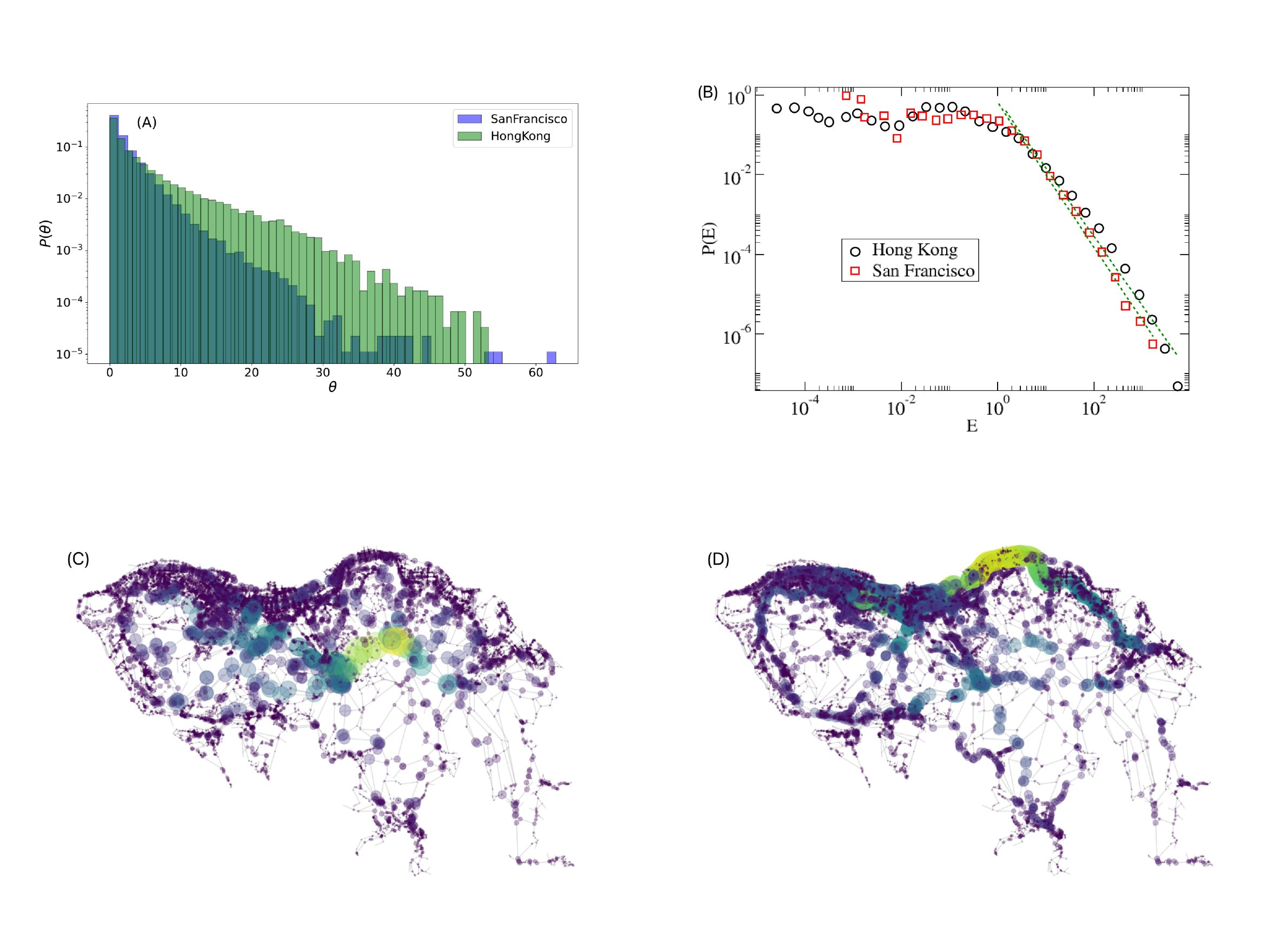}
	\caption{(A) Slope distribution for edges for the two networks of Hong Kong and San Francisco. (B) Excess effort distribution $P(E)$ for Hong Kong and San Francisco (computed for $500\times 500$ pairs of nodes). The dotted lines represent power law fits (with exponents $\gamma=1.72$ and $1.82$). (C,D) Distribution of the BC in Hong Kong (the size and color depend on the BC: the larger it is or the brighter it is and the larger the BC). On (C), we show the result computed on the 2d plane graph (neglecting the elevation) and on (D), we show the result when elevation is taken into account.}
	\label{fig:slope}
\end{figure*}

The slope distribution is also an interesting indicator of the intensity of elevation fluctuations experienced on a surfacic network. We show the result for the two different cities Hong Kong and San Francisco in Fig.~\ref{fig:slope}(top).

\subsection{Excess effort}

We computed the excess effort for the two cities and the results are shown in Fig.~\ref{fig:slope}(top, right). We observe that for $E>1$, the tail can reasonably be fitted by a power law of the form $P(E)\sim E^{-\gamma}$ with $\gamma=1.72$ for Hong Kong and $\gamma=1.82$ for San Francisco. These values are close to each other, but significantly different from the parabolic model for which we obtain $\gamma=2$. The occurrence of pairs of nodes for which the excess effort is very large is thus more likely 
in these two cities compared to the parabolic model. This is somehow expected as these real-world surfacic networks display more than one peak and are more rugged. The results are shown here for $500\times 500$ pairs of nodes randomly chosen. We plotted (see SI, Fig.~S6) the distribution for different sizes (up to $1000\times 1000$) which displays a quick convergence with size. 


\subsubsection{Betweenness centrality: spatial distribution}

As expected from the theoretical considerations and results obtained for toy models (see above for results on the parabolic model), when the elevation of a node is too large it becomes something to avoid. As a consequence, shortest paths avoid elevation peaks and low elevation nodes become more central than in a flat city.

This is confirmed in the specific case of Hong Kong where we show both the BC map (see Fig.~\ref{fig:slope}) in the 2d case (elevation is not taken into account) and the full 3d case (with elevation). The results confirm our theoretical analysis: in the 2d case, central nodes (in the spatial sense) are also the ones with the largest BC and when we take elevation into account we observe
that large BC nodes are on the boundary of the island (especially the northern one where the density of roads is larger).

\section{Discussion}

We have described a class of networks that has not been formally considered before, which we call surfacic network. For these networks, the shape of the surface governs the elevation fluctuations and therefore movement through the network, in our example pedestrian experience and behaviour. The difference between non-surfacic and surfacic version of a network is crucial in general for any network where there is a cost associated with an elevation difference, or indeed a benefit.

We illustrate using an urban pedestrian network, with our  results showing that elevation changes the network dynamics of city movement potential. This category of networks potentially represents a powerful framework for analyzing and modeling complex systems that are located on, or confined to, some kind of surface.

A more definitive and comprehensive characterization of supply networks is needed. While this paper has provided a minimalistic approach, a thorough mapping of surfacic network properties onto established graph-theoretic concepts would be valuable. A potential direction is linking geometric aspects, such as elevation-based surfaces \(z = f(x,y)\), with graph theoretical properties of networks. Specifically, surfacic netrworks require tools to measure flatness, effort, connectivity, and weighted distance on a surface where elevation is systematically related to network traversal, setting them apart from other networks where proximity in \(z\) is unrelated to traversal effort. Second, the impact of directionality in surfacic networks needs further exploration. Defining surfacic networks as directed graphs (digraphs) opens up a rich area of study. Pedestrian networks, for instance, may be navigated differently based on direction, and revisiting the current work with directionality in mind could reveal new insights. A digraph model could capture differences in path efficiency, excess effort, and weighted centrality depending on direction, prompting questions such as whether a lazy path uphill is equivalent to one downhill, and under what circumstances they differ.

Potentially, surfacic networks represent a versatile and interdisciplinary concept with applications across various fields. For example, they might be used to model chemical proceses on biological structures, such as trees, forests or coral reefs, where topography influences performance (e.g. by governing exposure to light). Surfacic network measures that we have defined, such as excess effort, would seem to have potential in measuring benefits and costs of traversing such networks, for example, in building physical infrastructure, or in expending energy when moving through a surfacic network; and benefits of capitalising on local slopes, such as when using gravity in hydrological systems engineering, or estimating profit potential of alternative paths in a network representation of a financial derivative instruments. By capturing explicit relationships between network topology and an underlying geometric topography, these networks provide valuable insights into the structure, behavior, and dynamics of complex systems situated on surfaces, paving the way for advances in science, engineering, and technology.

\section*{Acknowledgement}
We thank Mansur Boase for interesting remarks, and
MB thanks David Aldous for insightful comments on the MST on the random model.

\section{Data availability}

All the data used in this paper is publicly available at
\url{https://www.openstreetmap.org/} and at
\url{https://dataverse.harvard.edu/dataverse/global-urban-street-networks}.

The codes (in python) used for the analysis and visualization are available on the open repositery zenodo
at the address \url{https://zenodo.org/records/14557635} with the DOI: 10.5281/zenodo.14557635


\newpage

\section*{Supplementary Material}

\subsection*{Constructing the pedestrian network}

The steps for constructing the pedestrian networks are: build the
model with OSMnx using its built-in `walk' network type and without
any graph simplification. Then calculate this unsimplified graph's 2D
and 3D edge lengths. Then simplify the graph, summing length
attributes into the final simplified edges that represent street
segments between intersections/dead-ends.

There are very few degree-2 nodes in the graph (less than $0.1\%$ of
 nodes). These occur when the edges' OSM IDs change at that node
 point, suggesting different named streets  meeting at this intersection.

 We give below the pseudocode for the operations performed: downloading the network, adding elevation data,
processing edge attributes, simplifying the graph, and analyzing key
statistics.\\

\begin{algorithm}[H]
\caption{Modeling Pedestrian Networks with Elevation Data}
\SetAlgoLined

\textbf{Input:} Location string, Google API key\;
\textbf{Output:} Processed 3D pedestrian network stored as a GeoPackage\;

\textbf{Step 1: Download and preprocess network}\;
Download the walking network for the specified location using OSMnx\;
Add node elevations to the network using the Google API key\;

\textbf{Step 2: Compute additional edge attributes}\;
Project the network to a 2D coordinate system\;
\ForEach{edge $(u, v, k)$ in the network}{
    Compute 2D edge length using Euclidean distance\;
    Compute 3D edge length using Euclidean distance in 3D space\;
}

\textbf{Step 3: Simplify network}\;
Define aggregation rules for edge attributes (e.g., summation of lengths and travel times)\;
Simplify the network while retaining specified edge attributes\;
Add edge grades to the simplified network\;

\textbf{Step 4: Save the processed network}\;
Save the simplified network as a GeoPackage file\;

\textbf{Step 5: Analyze network statistics}\;
Extract nodes and edges as GeoDataFrames\;
Compute summary statistics for 3D/2D length ratios\;
Compute summary statistics for differences between 3D and 2D lengths\;
Compute the standard deviation of node elevations\;

\end{algorithm}

\subsection*{Steep versus smooth slope}

We consider the two possible paths from $A$ to $B$ shown in the
Fig.~S1.
\begin{figure}[h!]
	\centering
	\includegraphics[angle=0, width=0.40\textwidth]{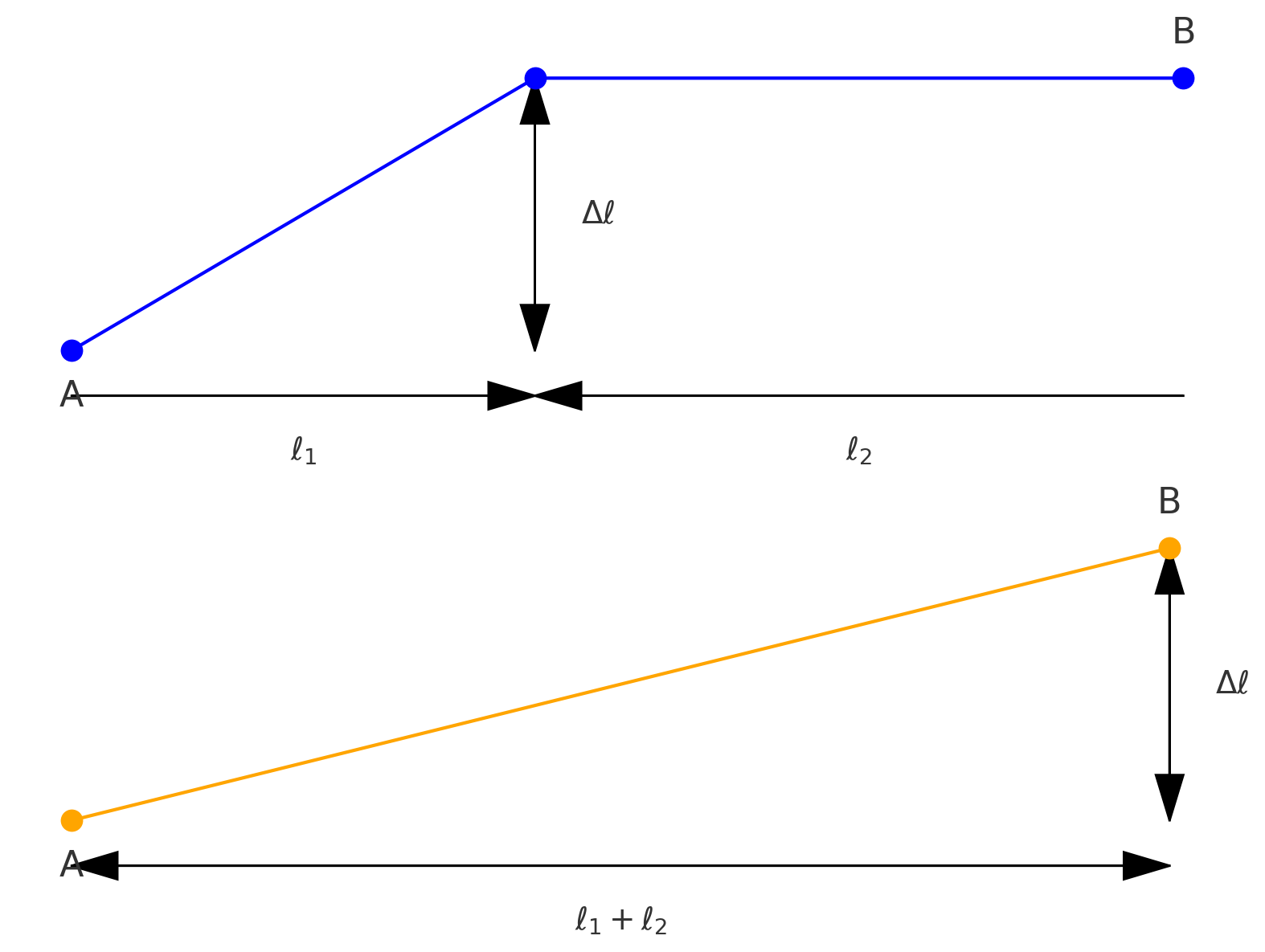}
	\caption{Figure S1. Two different paths from $A$ to $B$: (top)
          a steep slope followed by a flat path, (bottom) a smoother
          slope.  }
	\label{fig:example}
      \end{figure}
For the first path, the total weight is
\begin{align}
W_1=\sqrt{\ell_1^2+\Delta\ell^2}+\mu\Delta\ell +\ell_2
\end{align}
while for the smoother slope, we have
\begin{align}
  W_2=\sqrt{(\ell_1+\ell_2)^2+\Delta\ell^2}+\mu\Delta\ell
\end{align}
It is easy to check that we always have $W_1>W_2$ which implies that
the smoother slope path will always be preferred with this choice of
weight (one could for example introduce metabolic energy costs
considerations \cite{Kim}).

\subsection*{Effect of $\mu$ on the arduousness}

\begin{figure}[h!]
	\centering
	\includegraphics[angle=-90, width=0.50\textwidth]{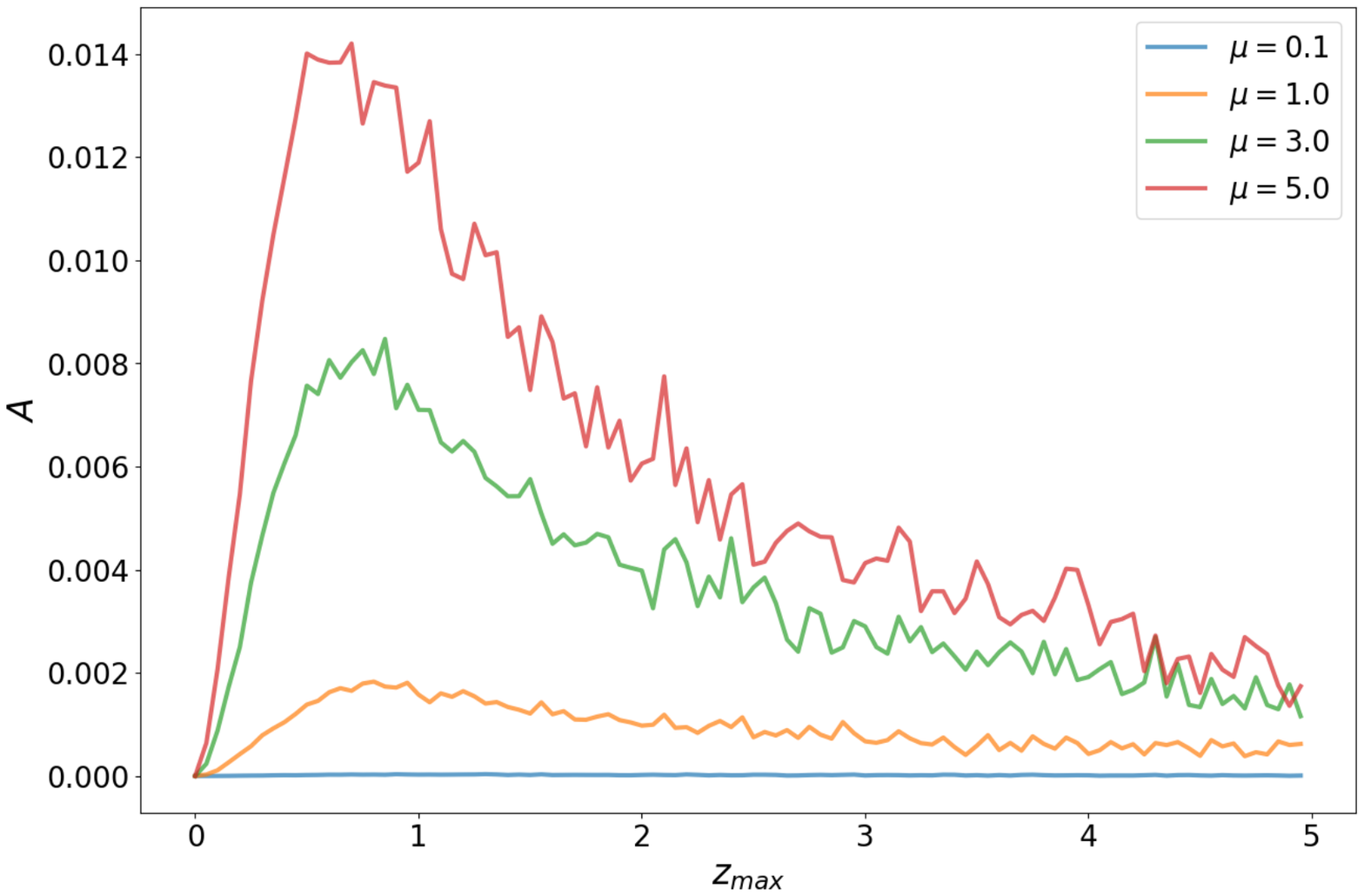}
	\caption{Figure S2. Arduousness vs. $z_{max}$ for the parabolic model
          with different values of $\mu$ (for $N=100$ and averaged
          over $100$ configurations). }
	\label{fig:mu}
\end{figure}

We show here the arduousness versus $z_{max}$ for the parabolic model
for different values of $\mu$ (Fig.~S\ref{fig:mu}).

The arduousness increases with $\mu$: for larger values of $\mu$,
optimal paths favor a small excess effort and not on the length of the
path, leading on average to a larger arduousness. The maximum value of
$A$ is obtained for $z_{max}\approx 1$.

\subsection*{The minimum spanning tree for the random model}

\begin{figure}[h!]
	\centering
	\includegraphics[width=0.30\textwidth]{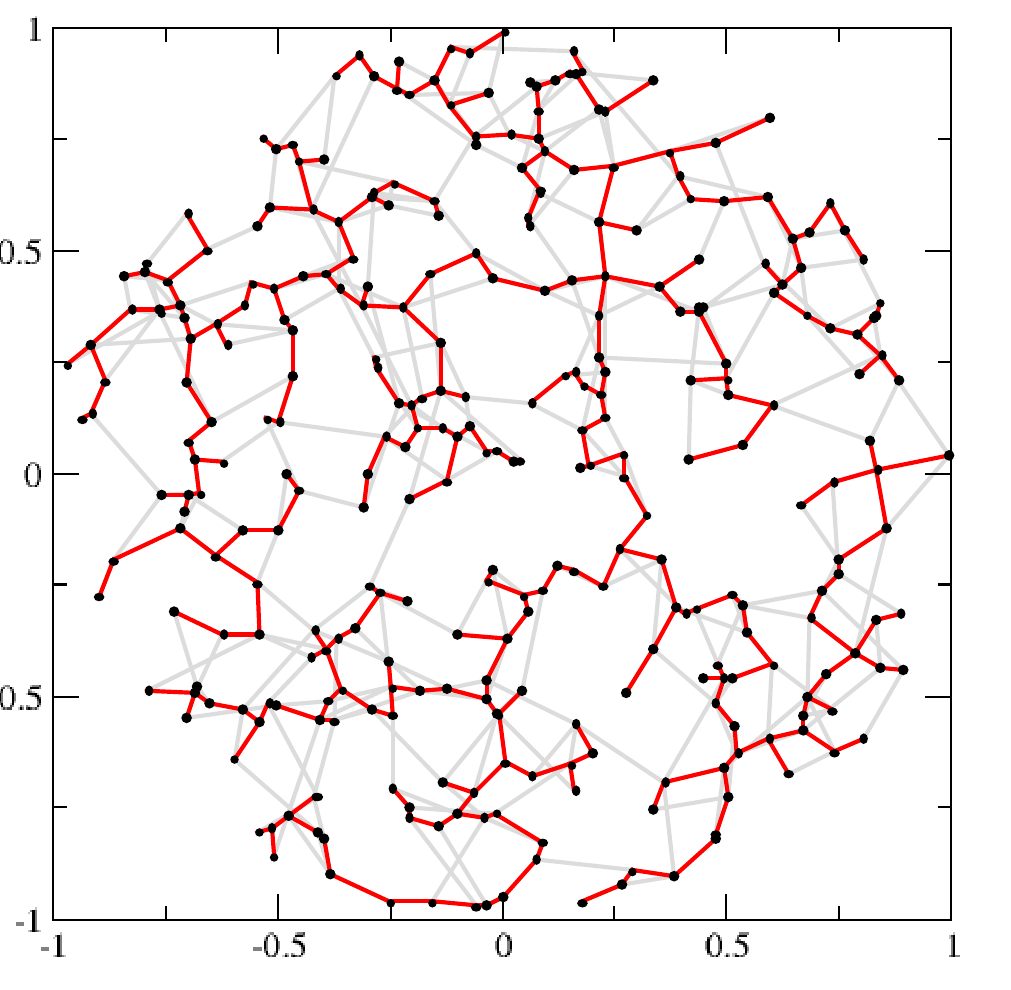}
	\caption{Figure S3. Minimum spanning tree for a null model (X-ray view
          from above). Comparison for
          $\gamma=0$ (gray links) and $\gamma=1$ (red links). }
	\label{fig:compare}
\end{figure}

A minimum spanning tree (MST) or minimum weight spanning tree is a
subset of the edges of a connected,
edge-weighted undirected graph that connects all the vertices
together, without any cycles and with
the minimum possible total edge weight. When the weight is the
euclidean distance, we obtain the Euclidean minimum spanning tree
(EMST). In order to illustrate the importance of elevation, we compute
for the same set of points, the EMST for the flat network ($\gamma=0$,
gray links) and for $\gamma=1$ (red links). Results shown in Fig.~S\ref{fig:compare}.
It is difficult to draw some conclusion from this small example, but
although the overall structure seems to be same
whether we take into account the elevation or not, there are some
important differences. Indeed, the links that
are used as bridges between different clusters of nodes links seem to
vary for the two minimum spanning trees. Further studies are certainly
needed to clarify this point.

\subsection*{Empirical results}

\subsubsection*{Elevation distribution}

\begin{figure}[ht!]
	\centering
	\includegraphics[width=0.5\textwidth]{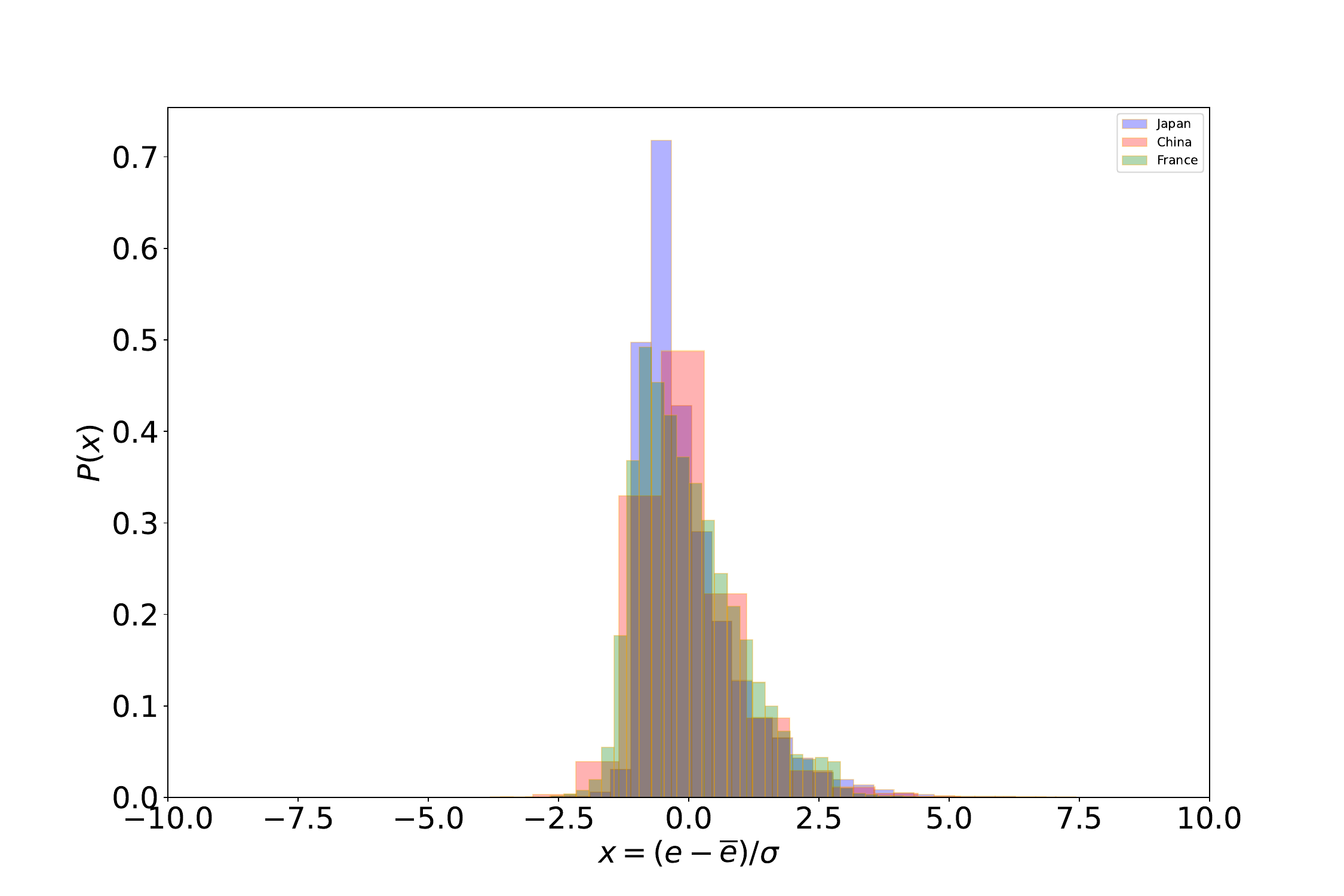}
	\caption{Figure S4. Probability distribution of the normalized elevation
          for different countries (USA, China, Japan, India).}
	\label{fig:Pxnorm}
\end{figure}

The first simple measure concerns elevation fluctuations in different
cities. We plot the distribution (Fig.~S\ref{fig:Pxnorm}) of the normalized elevation
$x=(e-\overline{e})/\sigma$ where $\overline{e}$ is the average for
each city (and $\sigma$ the corresponding std).  The cities belong to
different countries (USA, Japan, India, China) and the data comes from \cite{Boeing2017}.
More analysis is needed here, but despite the lack of apparent
universality, it seems that for all countries, this distribution is skewed (to the right).

\subsubsection*{Elevation versus Gini}

\begin{figure}[ht!]
	\centering
	\includegraphics[width=0.5\textwidth]{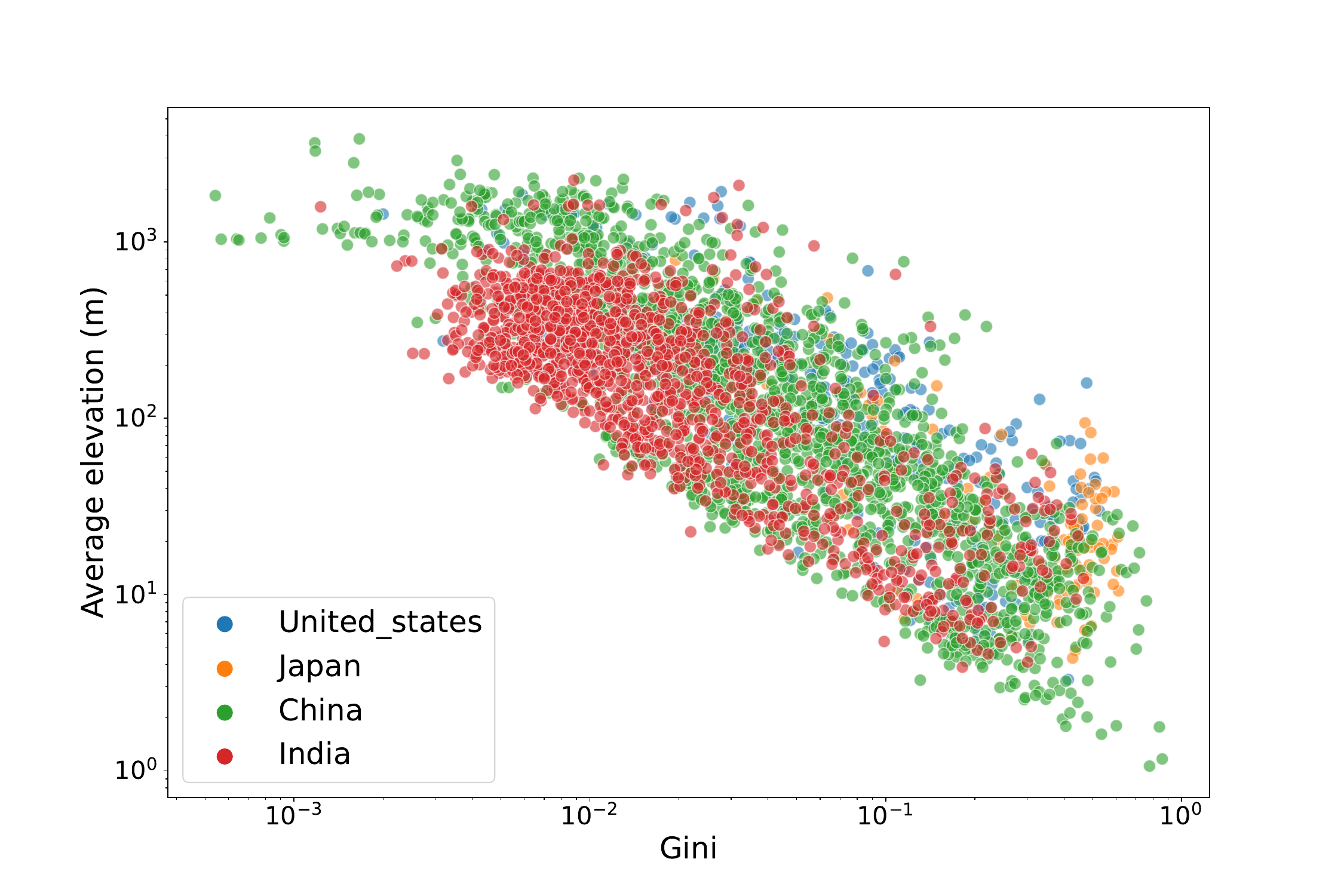}
	\caption{Figure S5. Average elevation for cities in various countries versus their Gini coefficient (data from \cite{Boeing2017}). }
	\label{fig:gini}
\end{figure}

For each city, we compute the average elevation of nodes
$\overline{z}$ and the Gini coefficient (see for example \cite{Dixon})
of elevations. In Fig.~S\ref{fig:gini}, we show this Gini coefficient
versus
the average elevation computed for cities in the USA, China, Japan,
and India.
We observe that the average elevation varies roughly as
$\overline{z}\sim 1/G$.

\subsubsection*{Effect of the number of pairs of nodes}
\begin{figure}
	\centering
	\includegraphics[width=0.5\textwidth]{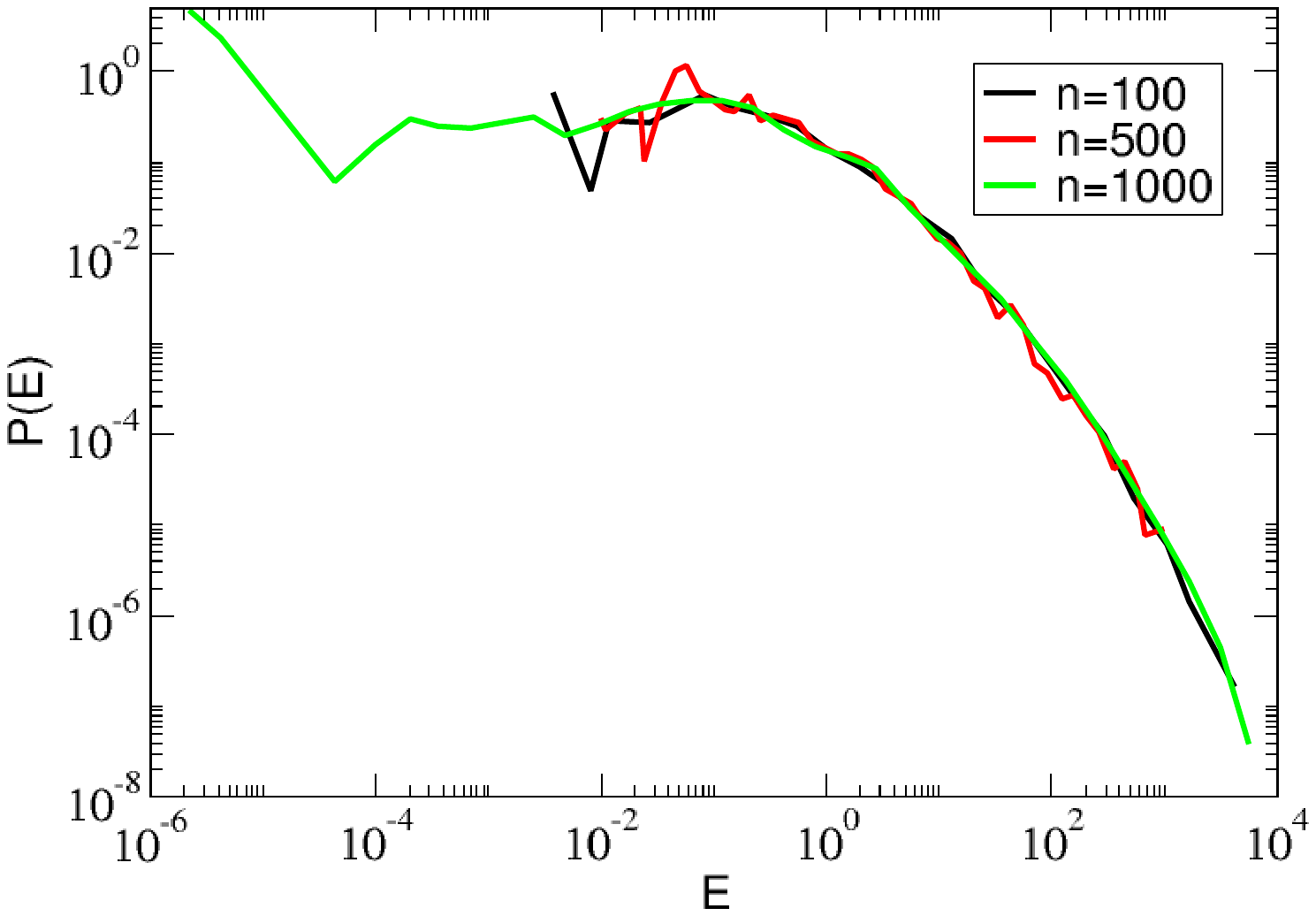}
	\caption{Figure S6. Effort probability distribution for different $n\times
          n$ pairs of nodes selected randomly. Results are shown here
          in the Hong Kong case. }
	\label{fig:cv}
      \end{figure}

We study here the effect of sampling of the pairs of nodes on the
statistics of effort. We select a number $n$ of randomly chosen nodes and compute
the effort for the $n\times n$ pairs of nodes. We compute the average
effort and its standard deviation and show the result in Fig.~S\ref{fig:cv}.

\end{document}